\documentclass[aps,twocolumn,showpacs,preprintnumbers,amsmath,amssymb,groupedaddress,superscriptaddress]{revtex4-1}

\usepackage{graphicx}
\usepackage{dcolumn}
\usepackage{bm}
\usepackage{xspace} % dynamic space in definitions
\usepackage{xcolor}

\begin{document}
\preprint{}

\title{IceCube search for dark matter annihilation in 
  nearby galaxies and galaxy clusters}

\affiliation{III. Physikalisches Institut, RWTH Aachen University, D-52056 Aachen, Germany}
\affiliation{School of Chemistry \& Physics, University of Adelaide, Adelaide SA, 5005 Australia}
\affiliation{Dept.~of Physics and Astronomy, University of Alaska Anchorage, 3211 Providence Dr., Anchorage, AK 99508, USA}
\affiliation{CTSPS, Clark-Atlanta University, Atlanta, GA 30314, USA}
\affiliation{School of Physics and Center for Relativistic Astrophysics, Georgia Institute of Technology, Atlanta, GA 30332, USA}
\affiliation{Dept.~of Physics, Southern University, Baton Rouge, LA 70813, USA}
\affiliation{Dept.~of Physics, University of California, Berkeley, CA 94720, USA}
\affiliation{Lawrence Berkeley National Laboratory, Berkeley, CA 94720, USA}
\affiliation{Institut f\"ur Physik, Humboldt-Universit\"at zu Berlin, D-12489 Berlin, Germany}
\affiliation{Fakult\"at f\"ur Physik \& Astronomie, Ruhr-Universit\"at Bochum, D-44780 Bochum, Germany}
\affiliation{Physikalisches Institut, Universit\"at Bonn, Nussallee 12, D-53115 Bonn, Germany}
\affiliation{Universit\'e Libre de Bruxelles, Science Faculty CP230, B-1050 Brussels, Belgium}
\affiliation{Vrije Universiteit Brussel, Dienst ELEM, B-1050 Brussels, Belgium}
\affiliation{Dept.~of Physics, Chiba University, Chiba 263-8522, Japan}
\affiliation{Dept.~of Physics and Astronomy, University of Canterbury, Private Bag 4800, Christchurch, New Zealand}
\affiliation{Dept.~of Physics, University of Maryland, College Park, MD 20742, USA}
\affiliation{Dept.~of Physics and Center for Cosmology and Astro-Particle Physics, Ohio State University, Columbus, OH 43210, USA}
\affiliation{Dept.~of Astronomy, Ohio State University, Columbus, OH 43210, USA}
\affiliation{Dept.~of Physics, TU Dortmund University, D-44221 Dortmund, Germany}
\affiliation{Dept.~of Physics, University of Alberta, Edmonton, Alberta, Canada T6G 2E1}
\affiliation{D\'epartement de physique nucl\'eaire et corpusculaire, Universit\'e de Gen\`eve, CH-1211 Gen\`eve, Switzerland}
\affiliation{Dept.~of Physics and Astronomy, University of Gent, B-9000 Gent, Belgium}
\affiliation{Dept.~of Physics and Astronomy, University of California, Irvine, CA 92697, USA}
\affiliation{Laboratory for High Energy Physics, \'Ecole Polytechnique F\'ed\'erale, CH-1015 Lausanne, Switzerland}
\affiliation{Dept.~of Physics and Astronomy, University of Kansas, Lawrence, KS 66045, USA}
\affiliation{Dept.~of Astronomy, University of Wisconsin, Madison, WI 53706, USA}
\affiliation{Dept.~of Physics and Wisconsin IceCube Particle Astrophysics Center, University of Wisconsin, Madison, WI 53706, USA}
\affiliation{Institute of Physics, University of Mainz, Staudinger Weg 7, D-55099 Mainz, Germany}
\affiliation{Universit\'e de Mons, 7000 Mons, Belgium}
\affiliation{T.U. Munich, D-85748 Garching, Germany}
\affiliation{Bartol Research Institute and Department of Physics and Astronomy, University of Delaware, Newark, DE 19716, USA}
\affiliation{Dept.~of Physics, University of Oxford, 1 Keble Road, Oxford OX1 3NP, UK}
\affiliation{Dept.~of Physics, University of Wisconsin, River Falls, WI 54022, USA}
\affiliation{Oskar Klein Centre and Dept.~of Physics, Stockholm University, SE-10691 Stockholm, Sweden}
\affiliation{Department of Physics and Astronomy, Stony Brook University, Stony Brook, NY 11794-3800, USA}
\affiliation{Department of Physics, Sungkyunkwan University, Suwon 440-746, Korea}
\affiliation{Dept.~of Physics and Astronomy, University of Alabama, Tuscaloosa, AL 35487, USA}
\affiliation{Dept.~of Astronomy and Astrophysics, Pennsylvania State University, University Park, PA 16802, USA}
\affiliation{Dept.~of Physics, Pennsylvania State University, University Park, PA 16802, USA}
\affiliation{Dept.~of Physics and Astronomy, Uppsala University, Box 516, S-75120 Uppsala, Sweden}
\affiliation{Dept.~of Physics, University of Wuppertal, D-42119 Wuppertal, Germany}
\affiliation{DESY, D-15735 Zeuthen, Germany}

\author{M.~G.~Aartsen}
\affiliation{School of Chemistry \& Physics, University of Adelaide, Adelaide SA, 5005 Australia}
\author{R.~Abbasi}
\affiliation{Dept.~of Physics and Wisconsin IceCube Particle Astrophysics Center, University of Wisconsin, Madison, WI 53706, USA}
\author{Y.~Abdou}
\affiliation{Dept.~of Physics and Astronomy, University of Gent, B-9000 Gent, Belgium}
\author{M.~Ackermann}
\affiliation{DESY, D-15735 Zeuthen, Germany}
\author{J.~Adams}
\affiliation{Dept.~of Physics and Astronomy, University of Canterbury, Private Bag 4800, Christchurch, New Zealand}
\author{J.~A.~Aguilar}
\affiliation{D\'epartement de physique nucl\'eaire et corpusculaire, Universit\'e de Gen\`eve, CH-1211 Gen\`eve, Switzerland}
\author{M.~Ahlers}
\affiliation{Dept.~of Physics and Wisconsin IceCube Particle Astrophysics Center, University of Wisconsin, Madison, WI 53706, USA}
\author{D.~Altmann}
\affiliation{Institut f\"ur Physik, Humboldt-Universit\"at zu Berlin, D-12489 Berlin, Germany}
\author{J.~Auffenberg}
\affiliation{Dept.~of Physics and Wisconsin IceCube Particle Astrophysics Center, University of Wisconsin, Madison, WI 53706, USA}
\author{X.~Bai}
\thanks{Present address: Physics Department, South Dakota School of Mines and Technology, Rapid City, SD 57701, USA}
\affiliation{Bartol Research Institute and Department of Physics and Astronomy, University of Delaware, Newark, DE 19716, USA}
% ...

\collaboration{IceCube Collaboration}
\noaffiliation

\begin{abstract}
We present the results of a first search for self-annihilating dark matter in nearby galaxies and galaxy clusters using a sample of high-energy neutrinos acquired in 339.8 days of live time during 2009/10 with the IceCube neutrino observatory in its 59-string configuration. The targets of interest include the Virgo and Coma galaxy clusters, the Andromeda galaxy, and several dwarf galaxies. 
We obtain upper limits on the cross section as a function of the weakly interacting massive particle mass between 300~GeV and 100~TeV  for the annihilation into $b \bar{b}$ , $W^+W^-$, $\tau^+\tau^-$, $\mu^+\mu^-$, and $\nu \bar{\nu}$. A limit derived for the Virgo cluster, when assuming a large effect from subhalos, challenges the weakly interacting massive particle interpretation of a recently observed GeV positron excess in cosmic rays.
\end{abstract}
\maketitle

%\linenumbers

\section{Introduction}
\label{sect:introduction}
There is compelling astronomical evidence for the existence of dark matter, although its nature remains unknown.
Among the theories providing suitable particulate candidates~\cite{bib:review1}, those that consider weakly interacting massive particles (WIMPs) are favored~\cite{bib:review2}. If stable particles exist with a mass between 10~GeV and multi-TeV that interact via the  electroweak force, they  would be produced and annihilate in thermal equilibrium in the early Universe. The cooling of the Universe would then naturally lead to a freeze-out with a relic density consistent with the measured dark matter abundance. %~\cite{bib:relic}. 
This annihilation process, producing Standard Model particles including neutrinos, is expected to take place in dark matter dense regions of the present Universe. %The annihilation process is expected to give rise to electrons and positrons, antiprotons, photons, and neutrinos that can be searched for.
 Promising sites for the observation of neutrinos from dark matter annihilation~\cite{bib:Strigari} include the cores of the Sun~\cite{bib:Sun} and Earth~\cite{bib:Earth}, as well as our Galactic halo~\cite{bib:Halo,bib:Halo2} and center~\cite{bib:GC}. 
Here we extend previous IceCube searches for self-annihilating Galactic dark matter to extra-Galactic sources. Potentially attractive targets are low surface brightness galaxies (also called dwarf spheroidals), clusters of galaxies, and large galaxies. 
%Here we examine for the first time the sensitivity of the IceCube detector and set limits to dark matter annihilation in close-by galaxies, galaxy clusters~\cite{bib:Clusters} and dwarf galaxies. 
Such searches have been  proposed~{\cite{bib:Proposed1,bib:Proposed2,bib:Proposed3} and are  also motivated by recent observations of a GeV positron excess seen by PAMELA~\cite{bib:Pamela} and confirmed by FERMI~\cite{bib:Fermi_positron} and recently by AMS-02~\cite{bib:AMS}. These positrons may originate from nearby astrophysical sources such as pulsars~\cite{bib:Pulsars}, but they could also be a hint for a leptophilic dark matter particle in the TeV mass range~\cite{bib:bergstroem,bib:Leptophilic1,bib:Leptophilic2}. The annihilation of such particles is expected to provide a flux of high-energy neutrinos that can be tested by neutrino experiments.

%N-body simulations of gravitational dark matter interactions~\cite{bib:aquarius,bib:vialactae,bib:Nbody} provide dark matter density distributions, $\rho(\vec{r})$, that suggest %the importance of dwarf galaxies and finer sub-clusters~\cite{bib:Subcluster}. 
%self-bound overdensities with masses down to $10^{-6} M_\odot$~\cite{bib:Gao}.
%These dense substructures lead to increased annihilation rates by several orders magnitude.
%to boost factors 
%that can increase the annihilation rates by several orders of magnitude.
 
%We examine for the first time sensitivities with IceCube for these objects and present limits.
%The results presented in this paper are used to constrain  $\left< \sigma_A v \right>$, the product of the self-annihilation cross section $\sigma_A$ and dark matter velocity $v$, averaged over the dark matter velocity distribution, as function of the dark matter particle mass $m_\chi $. \\

%
\begin{figure}
\includegraphics[width=0.9\columnwidth]{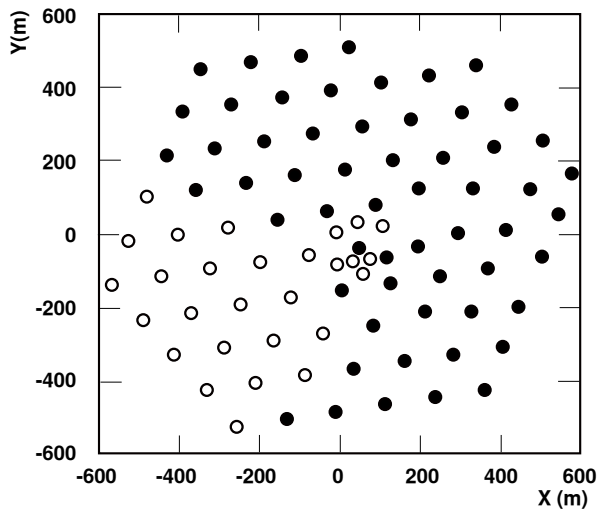}
	\caption{Schematic top view of the IceCube detector. The circles represent the positions of the 86 strings with 60 DOMs positioned at depths between 1450 and 2450~m; filled circles indicate the 59 strings with which the data for the analysis presented was obtained.}
	\label{fig:IceCubeDetector}
\end{figure}

%\section{Indirect dark matter search targets}
%Dwarf galaxies

The dark matter searches presented in this paper consider three types of astrophysical target objects: dwarf galaxies, the Andromeda spiral galaxy, and galaxy clusters.
Dwarf spheroidal galaxies are promising targets for indirect dark matter searches due to their estimated large dark matter densities, astrophysical simplicity, low luminosity, and absence of known background processes that could produce high-energy neutrinos. For a recent review, see Ref.~\cite{bib:Walker}. The detection sensitivity may be increased by stacking several target objects.
The dark matter halo of Andromeda (M31), the nearest spiral galaxy, is relatively well understood~\cite{bib:Andromeda}, with small uncertainties on the dark matter density profile.
Galaxy clusters (see, e.g., Ref.~\cite{bib:voit}) are the largest virialized objects observed in the Universe with $\approx$ 85\%, 12\%, and 3\% of the total mass provided by dark matter, intra\-cluster gas, and baryonic matter in galaxies \cite{bib:cluster_properties}, respectively. Their dark matter halo distribution appears to be well reproduced by N-body simulations for the  gravitational structure formation. 

N-body simulations of gravitational dark matter interactions~\cite{bib:aquarius,bib:vialactae,bib:Nbody} provide dark matter density distributions, $\rho(\vec{r})$,  that suggest self-bound overdensities. The minimal observed sizes are limited by the simulation resolution of about $10^5 M_\odot$. However, it has been suggested that much smaller protohalos may form. The minimal mass depends on the assumed decoupling temperature and may be in the range of  $10^{-11}$ to almost $10^{-3} M_\odot$~\cite{bib:cutoff}. In leptophilic models, the range may be extended to higher cutoff masses; see, e.g., Ref.~\cite{bib:cutoff2}.  In this analysis, we refer to a specific model assuming a cutoff mass of $10^{-6} M_\odot$~\cite{bib:Gao}.
%N-body simulations of gravitational dark matter interactions~\cite{bib:aquarius,bib:vialactae,bib:Nbody} provide dark matter density distributions, $\rho(\vec{r})$, that suggest %the importance of dwarf galaxies and finer sub-clusters~\cite{bib:Subcluster}. 
%self-bound overdensities with masses down to $10^{-6} M_\odot$~\cite{bib:Gao}.
These dense substructures would increase annihilation rates in galaxy clusters by several orders of magnitude.
%to boost factors 
%that can increase the annihilation rates by several orders of magnitude.

The results presented in this paper are used to constrain  $\left< \sigma_A v \right>$, the product of the self-annihilation cross section $\sigma_A$ and dark matter velocity $v$, averaged over the dark matter velocity distribution, as function of the dark matter particle mass $m_\chi $. \\

%Segue 1  is considered to be the least luminous galaxy found so far~\cite{bib:Geha}. 	
%Warum interessant, welche untersucht (Tabelle) [Draco, Coma-Berenices, Ursa Major II, Segue 1]\\
\begin{table*}
\centering
		\begin{tabular}{l | l | l | l | l| l |l }
		\hline
		Source           & Right              & Declination      & Distance & Mass               & log$_{10} J_{\rm NFW}$     & Boost factor   \\
		                 & ascension          &                  & [kpc]    & [M$_\odot$] & [GeV$^2$cm$^{-5}$] \\
		\hline
     Segue 1         & 10h 07m 04s  & +16$^\circ$04'55''  & 23    & 1.58$\times 10^{7}$ & 19.6 $\pm$ 0.5~\cite{bib:Jfactors}   & Not considered    \\
     Ursa Major II   & 08h 51m 30s  & +63$^\circ$07'48''  & 32    & 1.09$\times 10^{7}$ & 19.6 $\pm$ 0.4~\cite{bib:Jfactors}   & Not considered    \\
     Coma Berenices  & 12h 26m 59s  & +23$^\circ$54'15''  & 44    & 0.72$\times 10^{7}$ & 19.0 $\pm$ 0.4~\cite{bib:Jfactors}   & Not considered    \\
     Draco           & 17h 20m 12s & +57$^\circ$54'55''  & 80    & 1.87$\times 10^{7}$ & 18.8 $\pm$ 0.1~\cite{bib:Jfactors}    & Not considered   \\
     \hline
     Andromeda       & 00h 42m 44s & +41$^\circ$16'09'' & 778         & 6.9$\times 10^{11}$ & 19.2~\cite{bib:Andromeda}$^\ast$  & 66    \\
     \hline 
     Virgo cluster     & 12h 30m 49s  & +12$^\circ$23'28'' & 22300       & 6.9$\times 10^{14}$ & 18.2~\cite{bib:Han}$^\ast$     & 980 \\
     Coma cluster          & 12h 59m 49s  & +27$^\circ$58'50'' & 95000       & 1.3$\times 10^{15}$ & 17.1~\cite{bib:Han}$^\ast$ & 1300      \\
		\hline       
		\end{tabular}
	\caption{A list of potential astrophysical dark matter targets, their locations \cite{bib:SIMBAD}, distances, and masses \cite{bib:Kuhlen}, as well as $J_{\rm NFW}$ factors (see Sec.~\ref{sect:Signal}) considered in this paper. 
Boost factors for Andromeda, Coma, and Virgo are applied, when subclusters are taken into account. 
According to Ref.~\cite{bib:Charbonnier}, subclusters in dwarf galaxies do not usefully boost the signal.
For the extended Virgo cluster, M87 was used as the central position. %The $J$ factors of the dwarf galaxies and their logarithmic uncertainties are taken from~\cite{bib:Jfactors}. %The $J$ factor of the galaxy clusters Virgo and Coma are taken from~\cite{bib:Clusters} and for the M31 galaxy (Andromeda) the $J$ factor is calculated based on the profile given in~\cite{bib:Andromeda}. 
$^\ast$For Andromeda and the galaxy clusters, no uncertainties are available.}

	\label{tab:Dwarfsclusters}
\end{table*}

\section{Principle of Detection and the IceCube Telescope}
\label{sect:IceCube}
IceCube was designed to detect neutrinos of all flavors through Cherenkov light emission of secondary particles created in the interaction of a neutrino of energies above $\mathcal{O}(100)$~GeV with the surrounding ice or the nearby bedrock. 

A major challenge is the suppression of the cosmic ray background.
When high-energy cosmic rays hit air molecules in the upper Earth atmosphere, they initiate extended air showers that produce highly energetic pions and kaons and subsequently muons and neutrinos. %, with a spectrum proportional to $\approx E^{-3.7}$ or steeper.
% and a spectrum proportional to $\approx E^{-2.7}$ for the small and uncertain neutrino component from heavy meson decays in the atmosphere. 
Muons with energies exceeding 500~GeV reach the IceCube detector from above and dominate the detector event rate. Muons that would arrive from below are absorbed in the Earth. Muon neutrinos with energies less than 100~TeV, however, traverse the Earth with negligible absorption losses. Selecting tracks that enter from below the horizon therefore strongly suppresses cosmic ray muons, with the exception of tracks reconstructed in the wrong hemisphere. With tight cuts on the reconstruction quality, misreconstructed tracks are rejected, and the final data sample is dominated by the irreducible background of atmospheric neutrinos.
 
The construction of the IceCube neutrino observatory at the geographic South Pole was completed in December 2010. The detector instruments a volume of roughly one cubic kilometer of clear Antarctic ice~\cite{bib:Ice} with 5160 digital optical modules (DOMs)~\cite{bib:DOMs} at depths between 1450 and 2450~m. Each DOM contains a 25.3~cm diameter Hamamatsu R7081-02 photomultiplier tube~\cite{bib:Hamamatsu} connected to a waveform recording data acquisition circuit capable of resolving pulses with nanosecond precision and having a dynamic range reaching at least 250 photoelectrons per 10~ns. The observatory, depicted in Fig \ref{fig:IceCubeDetector} for the 2009/10 configuration, also %~\cite{bib:Detector} 
includes the densely instrumented DeepCore subdetector~\cite{bib:DeepCore} and the surface air shower array IceTop~\cite{bib:IceTop}. 
At that time, the detector was partially instrumented with 3540 DOMs, attached to 59 electrical cable bundles (strings) in the ice. Each string carries power and communication between each of the 60 DOMs and the surface data acquisition building.
 
To reduce the contribution from random noise hits, a local coincidence condition was enforced that requires the vertical neighbors of the triggered DOMs to register hits within 1~$\mu$s of each other. A multiplicity condition, requiring 8 DOMs to exceed their discriminator threshold within a 5~$\mu$s time window, served as the primary trigger for this analysis. The trigger rate in the 59-string configuration ranged from 1200 to 1500~Hz. The increased rate occurs during the austral summer as the probability of pions generated in cosmic ray air showers to decay rather than interact increases in the warmer and thinner atmosphere~\cite{bib:Tilav}. %A preselection at the South Pole reduces the data set to $\approx$ 65~GB/day, which is small enough to be sent via satellite and is subject to full event reconstruction in the offline processing. 
%
%
%Here we restrict ourselves to muon neutrinos, as their direction can be well estimated by reconstructing their long tracks at high energies. 

\section{Signal Expectations}
\label{sect:Signal}
The energy distribution of the expected neutrino flux depends on the branching ratio of the dark matter annihilation channels. This quantity is highly model dependent, and we therefore study different extremes of the possible annihilation channels and assume a branching ratio of 100\% for each of them in turn. We consider soft neutrino spectra produced by the annihilation into quarks ($b\bar{b}$), harder spectra as produced by $W^+W^-$, $\tau^+\tau^-$, and $\mu^+\mu^-$, and line spectra by annihilation into the $\nu \bar{\nu}$ final state. 
%Since the energy distribution of the annihilation products strongly depend on the WIMP mass and mix of decay channels,

The expected neutrino flux is given by
\begin{eqnarray}
%\frac{d\phi_\nu}{dE}=\frac{\left< \sigma_A v \right>}{2\cdot4 \pi m^2_\chi}\sum_F f_F \frac{dN_F}{dE_F}\times J(\Delta\Omega) \quad ,
\frac{d\phi_\nu}{dE}=\frac{\left< \sigma_A v \right>}{4 \pi\cdot 2 m^2_\chi} \frac{dN_\nu}{dE}\times J(\Delta\Omega) \quad ,
\label{eq:phi} 
\end{eqnarray}
where $\left< \sigma_A v \right>$ is the velocity averaged annihilation cross section, $m_\chi$ is the mass of the dark matter particle and
% $f_F$ is the fraction of annihilations which produce a final state $F$ and 
$dN_\nu/dE$ is the corresponding differential muon neutrino yield per annihilation. %after neutrino oscillation have been taken into account. 
We include neutrino flavor oscillations in the long baseline limit~\cite{bib:longbase}, since the neutrino flavor distribution at Earth will be mixed through vacuum oscillations. % as a result of different neutrino travel distances. 
The expected spectra at Earth were determined as described in Ref.~\cite{bib:Halo}.  

The flux is proportional to the integral over the square of the dark matter density,

\begin{eqnarray}
J(\Delta\Omega)& = & \int_{\Delta\Omega}d\Omega \int_{l.o.s.}\rho(l)^2 dl\quad ,
%& \cdot & \frac{\rho^2(\sqrt{R^2_{\rm sc}-2lR_{\rm sc}cos\psi{\Omega}+l^2})}{R_{\rm sc}\rho^2_{\rm sc}}\quad ,
\label{eq:J}
\end{eqnarray}
where $l$ is the coordinate along the line of sight of the observer toward the object.
%where $\psi$ is the angle between the observer and the object and $l_{\rm max}$ is the upper limit of the integration, defined as 
%\begin{eqnarray}
%l_{\rm max}=\sqrt{R^2-\sin^2\psi R^2}+R\cos\psi\quad .
%\label{eq:lmax}
%\end{eqnarray}

For a smooth parametrization of the dark matter halo, we refer to the NFW profile~\cite{bib:NFW}, where
\begin{eqnarray}
  \rho(r) & = & \frac{\rho_0}{\frac{r}{R_s}\left(1+\frac{r}{R_s}\right)^2} \quad .
\end{eqnarray}
Here $\rho_0$ and $R_s$ are the characteristic density and radius. If the field of view, $\Delta\Omega$, is large enough to cover the complete dark matter halo, then the resulting $J$ factor is $J_{\rm NFW}  = 4\pi\rho^2_0 R^3_s/3D^2$, where $D$ is the distance to the object (see Table~\ref{tab:Dwarfsclusters}). 
To facilitate the comparison with the Fermi result for dwarf galaxies~\cite{bib:Jfactors}, we used the same J factor values. The J factor for Draco in Ref.~\cite{bib:Charbonnier} is smaller but consistent within 2 sigma of the quoted uncertainties. Part of this difference is due to the choice of dark matter profile, which continues to be debated~\cite{bib:VeraCiro}.

%Recent simulations suggest the presence of dark matter substructures i.e. self-bound overdensities within the main halo of galaxies. 
Taking dark matter substructures in the halo into account, a stronger signal is expected, even at larger distances from the center (see Fig.~\ref{fig:extended}). 
We use the following proposed parametrization of this effect (Refs.~\cite{bib:Gao} and~\cite{bib:Han}) for the boost factor $b$  and the profile $j$:
\begin{eqnarray}
b(M_V) & = & \frac{J_{\rm sub\mbox{-}cluster}}{J_{\rm NFW}}=1.6\times 10^{-3}\left(\frac{M_V}{M_\odot}\right)^{0.39}\nonumber\\
j(r)   & = & \frac{16 b(M_V)J_{\rm NFW}}{\pi\ln(17)}\frac{1}{r_V^2+16r^2}\quad  {\rm for}\quad r < r_V \nonumber\\
       & = & j(r_V){\rm e}^{-2.377(r/r_V-1)} \quad  {\rm for}\quad  r > r_V \quad .
\end{eqnarray}
Here $j$ is the line of sight integral over the squared density, and $J$ is the total integral, given by $J =\int_{\Delta\Omega}{j d\Omega}$. $M_V$ and $r_V$ denote the virial mass and radius of the halo. % within which the average density is 200 times the critical density of the Universe. 
For this optimistic parametrization, which allows for subhalo masses down to $10^{-6} M_\odot$, the effect of subhalos is largest for galaxy clusters, with boost factors of 1300 and 980 for the Coma and Virgo clusters, respectively, followed by Andromeda with a boost factor of 66 (see also Fig.~\ref{fig:extended}) and a boost factor close to 1 for dwarf galaxies~\cite{bib:Charbonnier}. The subclustering and corresponding boost factors is an active area of research, and  our results with and without subclusters likely bracket the probable range.

%Neutrino spectra take from Halo paper\\

%
\section{Data Selection}
\label{sect:Selection}
%Still the million times larger flux of down going muons remains a challenging background, 
Downward-going cosmic ray muons, which are detected $\mathcal{O}(10^6)$ times more frequently than atmospheric neutrinos, constitute the primary background for this analysis, even if only a small fraction of the events is misreconstructed as upward going. 
A series of event selections and higher-level event reconstructions were applied to remove these background events, while 
retaining upward-going tracks from muons induced by Earth-crossing neutrinos. %Such events were tagged at the South Pole at 
By this online filter, the rate was reduced to $\approx$ 35~Hz, and the events were transmitted %with other preselected events 
via satellite to the Northern Hemisphere where additional fits were applied offline.
%offline outside of the South Pole. 
Below we describe the reconstruction algorithms for the muon direction from the pattern of registered Cherenkov light as well as quality parameters used in this analysis. Toward the end of the section, we summarize the precuts and the final data selection. 
%A series of likelihood fits, which utilize the arrival time distribution of Cherenkov photons in each DOM assuming the expectation for an infinite muon track, was performed. 

First, reconstructions~\cite{bib:Reconstruction} were performed using a single photoelectron (SPE) likelihood, which uses the arrival time of the first Cherenkov photon hitting each DOM. The likelihood fit, initialized with a line-fit seed (LF), was later iterated eight times with random starting values to find the global optimum (SPE8). A multiple photoelectron (MPE) fit, which uses the likelihood description of the arrival time of the first least scattered Cherenkov photon in each DOM, given $N$ measured photons in that DOM, was then applied. Note that the MPE fit provides improved directional resolution for neutrinos at higher energies, while the SPE fit reconstruction is more efficient at rejecting events caused by bundles of cosmic ray muons. As a measure of the quality of the fit, the so-called reduced log likelihood $RLogL =-\ln \mathcal{L_{\rm MPE}}/(N_{\rm CH}-a)$  was calculated, where $N_{\rm CH}$ is the number of DOMs that registered a hit. This test variable is motivated by the relation $-\ln \mathcal{L}=\chi^2/2\, -\, c$ for normal distributions. The  constant $a$ was chosen ($a=2.5$ for $RLogL_{\rm MPE}$ and $a=2.0$ for $RLogL_{\rm SPE}$) such that $\left< RLogL \right>$ was approximately independendent of $N_{\rm CH}$. 

A substantial fraction of the atmospheric muon background results from two or more muons being produced in uncorrelated cosmic ray showers that enter the IceCube detector in one trigger window. To reduce this background, two muon tracks were reconstructed for each event after splitting the triggered DOMs, either in geometry or in time, into two groups. Each group of DOMs is used to reconstruct a track assuming a single muon hypothesis, resulting in two reconstructed muon tracks. Multiple muon tracks can alternatively be identified by grouping topologically connected hits both in time and in space.
A Bayesian likelihood ratio, $R_{\rm Bayes}$, compares the hypothesis of the upward-going muon track (SPE8) with the alternative hypothesis of a
down-going muon track, employing a likelihood that strongly suppresses upward-going directions.
%as the product of the standard likelihood and a Bayesian prior, which is based on the known zenith dependence of the down-going muon flux and the relative rates of atmospheric muons and neutrinos. 
%Requiring the zenith angle from both reconstructed tracks to traverse through the Earth reduces the coincident atmospheric muon
%background.

Minimally scattered Cherenkov photons were selected by defining a time window ranging from -15 to 75~ns between the expected arrival time from the reconstructed muon track and the first registered hit. The number of DOMs with such a direct hit, $N_{\rm direct}$, and the largest distance between them along the track, $L_{\rm direct}$, are measures of the track accuracy. The zenith and azimuth resolution $\sigma_{\rm cr}(\theta)$ and $\sigma_{\rm cr}(\phi)$ were determined from the Fisher information matrix, exploiting the Cramer Rao inequality, by using the set of hit DOMs and the corresponding average time delays due to light scattering.%, as well as expected zenith, $\sigma_{\rm cr}(\theta)$, and azimuth, $\sigma_{\rm cr}(\phi)$.

A set of precuts was introduced to reduce the data sample for the analysis.
 Only upward-going events, defined in the online filter to have fulfilled the requirements $N_{\rm CH} \ge 10$  and for the line fit zenith angle $\theta_{\rm LF} > 70^\circ$, followed by a successful SPE likelihood reconstruction  with the zenith angle of the SPE fit $\theta_{\rm SPE}>80^\circ$ and $RLogL_{\rm SPE}<8.2$, %$\ln \mathcal{L_{\rm SPE}}/(N_{\rm CH}-2.0)\le 8.2$
 were considered. The zenith angles of the tracks, reconstructed from the temporally and geometrically split subevents, 
%The zenith track angles of the time and geometrically split sub-events 
were not permitted to be $<57.3^\circ$. The zenith angle of the largest topological trigger split subevent was required to be larger than 80$^\circ$. All unsplit events were kept. To reduce the fraction of events with poorly reconstructed direction, only events with hit DOMs in  more than one string were kept, and initial loose cuts,  $(L_{\rm direct}/60 {\rm m})^2 + (N_{\rm direct}/15)^2 > 1$ and $\sigma_{\rm cr}(\theta)< 57.3^\circ$, were imposed.

Events surviving the precuts described above, predominantly misreconstructed atmospheric muons, were analyzed with a boosted decision tree (BDT)~\cite{bib:BDT}, a multivariate machine learning algorithm that was optimized to discriminate the neutrino signal and the atmospheric muon background. The BDT was trained on a background dominated data set at a low cut level and on a simulated signal from WIMPs annihilating into the $\tau^+\tau^-$ final state. The energy spectra for the simulation were obtained with DarkSUSY~\cite{darksusy}. 
To accommodate a broad range of WIMP masses, an average of the neutrino spectra between 300~GeV and 100~TeV was used. 

The following five event observables were found to offer the highest discriminating power between signal and background and were subsequently used as input to the BDT: $\log_{10}\left(\sqrt{\sigma_{\rm cr}(\theta)\cdot\sigma_{\rm cr}(\phi)}\right)$; 
the spatial angle between the MPE and line fits; $RLogL_{\rm MPE}$, $R_{\rm Bayes}$  and $L_{\rm direct}$. As an example, Fig.~\ref{fig:BDTtau} presents a comparison of the data with simulated atmospheric neutrinos events as a function of the BDT output variable. %For this plot, the BDT was trained to optimally separate the dark matter annihilations into $\tau^+\tau^-$ from the background. 
Table~\ref{tab:CutLevels} shows the corresponding data rates. The final data sample is dominated by atmospheric neutrinos. 
%Typical cut values are around 0.25.
\begin{figure}
\includegraphics[width=1.1\columnwidth]{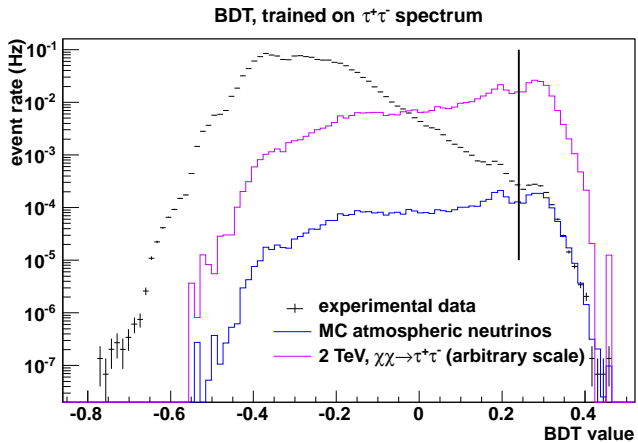}
  \caption{Comparison of the BDT value for all events passing the precuts compared to results from an atmospheric neutrino Monte Carlo. To illustrate the seperation power, the spectrum for 2~TeV WIMPs annihilating into $\tau^+\tau^-$ with arbitrary normalization, is also shown. The vertical line represents a typical cut value. In Table~\ref{tab:CutLevels} the corresponding data rates are shown.}
	\label{fig:BDTtau}
\end{figure}

\section{Data Analysis}
%Background determination\\
After applying the cut on the BDT output variable, the background was estimated from the data in a 5 deg wide zenith band centered around the nominal zenith positions of the sources. The statistical uncertainty primarily depends on the zenith position and ranges from  2.2\% for the Virgo Cluster to 4.2\% at the zenith position of Ursa Major II (for the $W^+W^-$ channel and 5 TeV WIMP mass assuming the NFW profile, see also Table~\ref{tab:Dwarfsclusters}).   
%Optimization\\
To define the selection criteria before analyzing the complete data set, both the cut value on the BDT and the search radius, defined as the maximal space angle between the nominal source position and the measured direction (using MPEFit), were simultaneously optimized. This was done by minimizing the quantity $\mu_{90}/\epsilon \propto \phi_\nu$, the average expected upper limit divided by the signal efficiency $\epsilon$.
Typical cuts for the BDT output value ranged between 0.08 for very soft spectra and 0.3 for very hard spectra. This optimization was 
performed for a WIMP mass of 5 TeV and was subsequently used for all assumed masses between 0.3 TeV and 100 TeV. Note that the selected mass values are indicated as dots in Fig.~\ref{fig:comparison_tau}.
As a cross-check, individual optimizations for all mass values were tested. This procedure was not followed because it would have increased the number of trials, and the cut criteria turned out to be rather similar.
%initially performed for all masses and annihilation channels individually. In order to minimize the number of trials, and because the cut criteria turned out to be rather similar, it was later decided to use the optimization obtained for a WIMP mass of 5 TeV for all assumed masses between 0.3 TeV and 100 TeV. 

%Extended Sources\\
The angular resolution of the IceCube detector for $\nu_\mu$ charged current events depends on the energy spectrum and thus on the WIMP mass and annihilation channel and is on the order of a few degrees. The corresponding point spread function was determined from simulated events for every assumed signal spectrum. Convolving the point spread function with the much narrower assumed NFW profile for the source did not change the functional shape significantly.
%After convolution with NFW profiles, the halos of dwarf galaxies are still sufficiently cusped such that the angular extension is smaller than the detector resolution. 
As discussed above, subclustering is important for the Andromeda galaxy and the Virgo and Coma galaxy clusters, leading to extended signal regions, as seen in Fig.~\ref{fig:extended}. In this case it is important to convolve the signal profile with the point spread function to estimate the signal efficiency.
%This is not the case if the effect of sub-clusters is included, as seen in Fig.~\ref{fig:extended} for the Andromeda galaxy and the Virgo and Coma galaxy clusters, leading to larger search bin sizes.

\begin{figure}
\includegraphics[width=1.1\columnwidth]{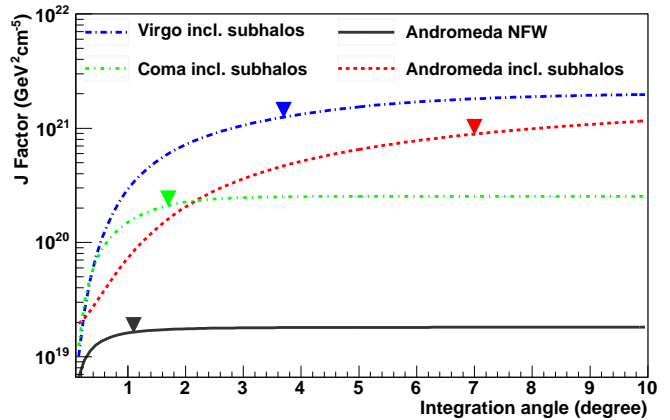}
	\caption{Cumulative distributions for the $J$ factors calculated assuming  subhalos, for Andromeda and the Coma and Virgo galaxy clusters. For comparison, the distribution for Andromeda assuming an NFW profile is also shown. The effects of both boosting and the widening of the distribution due to dark matter accumulations far from the center of the galaxies and galaxy clusters are visible. The search angle cuts for the $W^+W^-$ annihilation channel and 5 TeV WIMP mass are indicated by arrows. }
	\label{fig:extended}
\end{figure} 

%Stacking\\
To enhance the sensitivity, we investigated the stacking of several dwarf galaxies  
 by probing the corresponding signal regions simultaneously.
For each combination, the search radius and the BDT cut value were optimized as discussed above. To determine the combined flux, the $J$ factors of Table~\ref{tab:Dwarfsclusters}  were assumed.
% weighted by the $J$ factors of the sources. 
The best sensitivity was found by stacking Segue~1 and Ursa Major II, the sources with the strongest expected signal. A stacking of galaxy clusters has not yet been attempted.

\begin{table}
  \centering
		\begin{tabular}{l | l | l | l }
		\hline
		Cut        & Data      				& Atmospheric        & Atmospheric       \\% & Signal          \\
		level      & rate [Hz] 				& $\mu$ rate [Hz]         & $\nu$ rate [Hz]        \\% & eff. [\%]       \\
		\hline
%		filtered   & 35           			&                   &                  \\% &      100        \\
		Before BDT & 1.4          			&     0.92          &$4.9\times 10^{-3}$\\% &                 \\
		After BDT  & $1.4\times 10^{-3}$                 &$2.6\times 10^{-4}$ &$1.1\times 10^{-3}$                  \\
		\hline       
		\end{tabular}
	        \caption{Data, atmospheric muon, and neutrino expected background rates %for different cut progressions on the online filtered data sample ($\approx 35$ Hz).
before and after a typical cut on the BDT output value. The online filter rate is $\approx 35$ Hz.
 The Monte Carlo rates for atmospheric neutrinos and muons are meant to illustrate the background sources and are not used in the analysis.}%The signal efficiency for detecting neutrinos from Virgo in the $W^+W^-$ annihilation channel at 5 TeV relative to the filter level at the South Pole is shown for comparison.}
	\label{tab:CutLevels}
\end{table}

\section{Systematic Uncertainties and discussion of astrophysical uncertainties}
\label{sect:Systematics}
\begin{figure}
  \includegraphics[width=1.05\columnwidth]{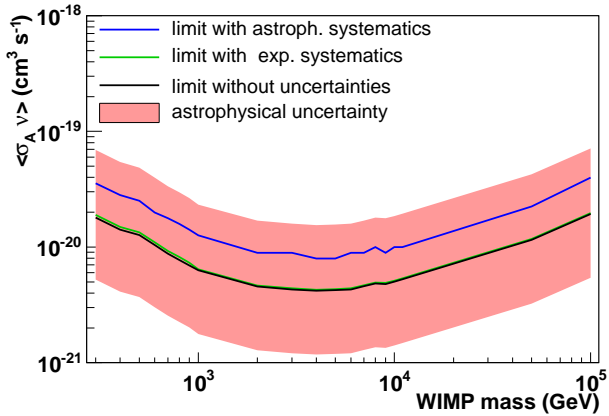}
  \caption{Impact of including uncertainties on the limit calculation for Segue~1 for annihilation into $W^+W^-$. For all other limits, the astrophysical uncertainties are not included to allow for inclusion of better estimates of the $J$ factors in the future.}
  \label{fig:systematics}
\end{figure}

\begin{table}
  \centering
  \begin{tabular}{l | c | c | r  }
    \hline
    Source                       & \multicolumn{3}{|c}{WIMP Masses (TeV)}\\
                     & $<$ 1  & 1 - 10   & $>$ 10  \\
    \hline
    %$\nu$ oscillations           & 6       &  6      & 6       \\
    %$\nu$ -nucleon cross-section & 7       & 5.5     & 3.5     \\
    %$\nu$ -propagation in ice    & <1      & <1      & <1      \\
    %Time, position calibration   & 5       & 5       & 5       \\
    %DOM sensitivity spread       & 6       & 3       & 10      \\
    Photon propagation in ice    & 20\%      & 20\%      & 15\%       \\
    Absolute DOM efficiency      & 15\%      & 10\%      &  5\%      \\
    \hline
    Total uncertainty            & 25\%      & 22\%      & 16\%      \\
    \hline       
  \end{tabular}
  \caption{Relative uncertainties of the dominating experimental systematics affecting the flux determination. The uncertainties were added in quadrature.
    %Relative decrease in percent of the sensitivity when incorporating systematic uncertainties.{\bf numbers still from Solar WIMP paper.}
  }
    \label{tab:Systematics}
\end{table}

By design, the comparison of events in the on- and off-source regions enables one to determine the background directly from the data. This eliminates most detector related systematic effects for the background estimation. 
The primary systematic uncertainties on the analysis are due to signal acceptance. In addition we discuss the impact of astrophysical uncertainties on our result. 
% in the modeling of the $\overline{J(\Delta\Omega)}$ factors 
% from the my, statistical uncertainties due to Monte Carlo statistics and the available off-source data for the background estimation.  

The astrophysical uncertainties mainly arise from the assumed dark matter densities and  profiles, entering the calculation as $J(\Delta\Omega)$, as well as from the scale of the dark matter substructures. While the uncertainties of the latter are difficult to assess, we list the uncertainties of $\log_{10}(J)$~\cite{bib:Jfactors} in Table~\ref{tab:Dwarfsclusters}. 

% \item {\bf Signal acceptance uncertainties}
The signal acceptance uncertainty is dominated by uncertainties in the ice properties and limitations in the detector simulation. Theoretical uncertainties, including muon propagation, the neutrino cross section, and the presence of the bedrock, each of
which have been studied in previous analyses (see, e.g., Ref.~\cite{bib:Systematics}) add approximately 6\% to the total uncertainty. %, are much smaller and on the order of 1\% or less. 
The uncertainty due to Monte Carlo simulation statistics and detector exposure as well as the individual track pointing uncertainty is much smaller.

\begin{figure}
\includegraphics[bb=0 0 567 384,width=1.05\columnwidth]{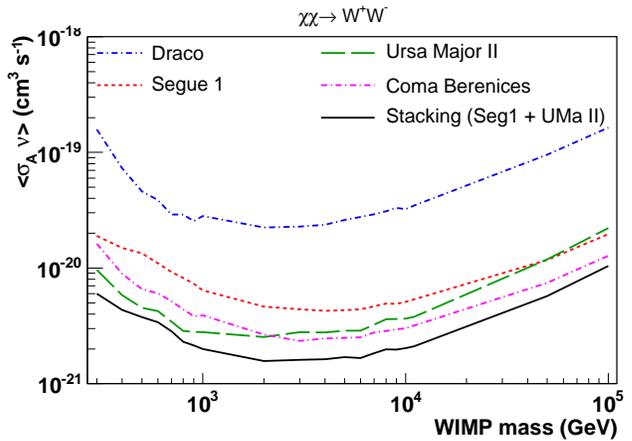}
\caption{%Comparison of sensitivities (dashed lines) and upper limits (solid lines) for Draco (blue), Segue 1 (red), Ursa Major II (green), Coma Berenices (violet) as well as for the stacking for Segue 1 and Ursa Major II (black) for decays into $W^+W^-$. 
  Upper limits for the annihilation into the $W^+W^-$ channel for the dwarf galaxies Draco, Segue~1, Ursa Major II, and Coma Berenices. Included is the stacking for Segue~1 and Ursa Major II.
%  The corresponding sensitivities differ by constant factors of 
The sensitivity curves (not shown) differ from the limit curves by multiplicative factors of
  0.58 (Segue~1), 
  0.90 (Ursa Major II), 
  0.61 (Draco),
  5.1 (Coma), and 
  1.0 (stacking).
}
\label{fig:dwarf}
\end{figure}

To assess the uncertainties on the ice properties, two ice models tuned to \emph{in situ} measurements with artificial light sources  were compared~\cite{bib:Spice},
% (SpiceMie, Spice1)
 and the ratio of the calculated sensitivities in both models was investigated as a function of WIMP mass and source direction. The observed discrepancy  between the models, also seen in the data/Monte Carlo comparison,  
%is largest for events near the horizon. The estimate for this systematic uncertainty in signal acceptance 
ranges between 10\% for tracks traversing the detector parallel to the strings and 20\% for larger zenith angles.%.(xxThe limited knowledge of ice properties as a function of depth and limitations in the detector simulation, is expected to produce an observed discrepancy between data and simulation for events near the horizon [36]. For nearly horizontal tracks the disagreement is maximal, with 30% more
%events observed in data compared to simulation predictions. Since we use the data itself to predict the number
%Concerning the signal model uncertainties, 

To assess the DOM sensitivity uncertainties, three Monte Carlo samples, with 90\%, 100\%, and 110\% of the nominal DOM sensitivity, were investigated as a function of WIMP mass and source direction. The observed discrepancy between the models is largest for low-energy events; see Table~\ref{tab:Systematics}. The estimate for this systematic uncertainty in signal acceptance ranges between 15\% for WIMP masses of 1~TeV and 5\% above 10~TeV.  

% The zenith dependent uncertainty of the background determination is dominated by the available off-source data and is yy\% for sources close to the horizon and zz\% for ..

\section{Results}
\label{sect:Results}
\begin{table*}
\centering
		\begin{tabular}{l | l | l | l | l | l | l | l | l | l | l }
		\hline
%                Source     & est. bg $\tau^+\tau^-$   & est. bg $b\bar{b}$ & est. bg $W^+W^-$ & est. bg $\mu^+\mu^-$  & est. bg $\nu\bar{\nu}$   \\
%                           & / events seen           & / events seen       & / events seen   & / events seen         & / events seen    \\
                Source     &   \multicolumn{2}{|c|}{  $\tau^+\tau^- $}      &  \multicolumn{2}{|c|}{ $b\overline{b}$} &  \multicolumn{2}{|c|}{ $W^+W^-$} &  \multicolumn{2}{|c|}{ $\mu^+\mu^-$}  &  \multicolumn{2}{|c}{ $\nu\bar{\nu}$}   \\
                     & \footnotesize{estimated} & \footnotesize{observed}    &  \footnotesize{estimated} & \footnotesize{observed}    & \footnotesize{estimated} & \footnotesize{observed}   & \footnotesize{estimated} & \footnotesize{observed} & \footnotesize{estimated} & \footnotesize{observed} \\
                     & \footnotesize{backgr.} & \footnotesize{events}    &  \footnotesize{backgr.} & \footnotesize{events}    & \footnotesize{backgr.} & \footnotesize{events}   & \footnotesize{backgr.} & \footnotesize{events} & \footnotesize{backgr.} & \footnotesize{events}    \\
		\hline
                Segue 1        & 8.7 & 10	     & 13.3 & 18	  & 8.2 & 12         & 8.7 &  10             & 4.3 & 6      \\
                Ursa Major II  & 7.4 & 8             & 5.2  & 1           & 7.4 & 8	     & 4.6 &  1	             & 3.5 & 1      \\
                Coma Berenices & 4.7 & 1	     & 11.6 & 4           & 4.7 & 1	     & 8.3 &  3	             & 4.7 & 1      \\
                Draco	       & 5.6 & 8	     & 13.4 & 15          & 5.6 & 8	     & 5.6 &  8	             & 4.5 & 8      \\
                Stacking (Seg1 + UMa II)&9.5 & 8     & 20.0 & 23          & 12.8 & 13	     & 9.5 &  8              & 5.3 & 4      \\
		\hline       
%		\end{tabular}
%	\caption{Dwarf estimated backgrounds and events found. The last line gives the results for the stacking of Segue 1 and Ursa Major II.}
%	\label{tab:Dwarfsresults}
%\end{table*}
%
%\begin{table*}
%\centering
%		\begin{tabular}{l | l | l | l | l | l}
%		\hline
%                source     & est. bg $\tau^+\tau^-$   & est. bg $b\bar{b}$ & est. bg $W^+W^-$ & est. bg $\mu^+\mu^-$  & est. bg $\nu_\mu\bar{\nu_\mu}$   \\
%                           & / events seen           & / events seen      & / events seen    & / events seen         & / events seen    \\
                Virgo (subhalos) & 92.1 & 89        & 322 & 325        & 103 & 102       & 92.1 & 89             & 94.7 & 92     \\
                Virgo (NFW)       & 9.6 & 9	     & 23.9 & 19          & 9.6 & 9         & 9.6 & 9              & 5.9 & 5    \\
                Coma (subhalos) & 17.5 & 17        & 35.8 & 40          & 14.0 & 15        & 14.0 & 15             & 13.5 & 15    \\
                Coma (NFW)        & 5.9 & 6	     & 13.7 & 13          & 5.9 & 6          & 5.9 & 6               & 4.8 & 5      \\
                \hline
                Andromeda (subhalos)  & 201 & 194	     & 413 & 418        & 201 & 194	     & 201 & 194           & 201 & 194    \\
                Andromeda (NFW)         & 6.4 & 2          & 6.7 & 1           & 6.4 & 2          & 6.4 & 2               & 4.3 & 0      \\
	\hline       
		\end{tabular}
	\caption{Number of events as estimated from background and as observed in the data, for dwarf galaxies, galaxy clusters, and Andromeda. In some cases the same cut values and bin sizes were used for different annihilation channels, leading to the same number of events.}
	\label{tab:Clusterresults}
\end{table*}

With the exception of cross-checks on small subsets of the data, the analysis was performed in a blind way: the signal optimization was done entirely on simulations, and the whole data set with full directional information was examined only after the selection criteria were finalized.

%. In this way a blind analysis was executed where the entire data set was examined only after the selection criteria were finalized.
%{\em minimizations of trial factors?}.
No significant excess beyond the background expectation was found.
Upper limits at the 90\% confidence level were calculated from the event and background numbers, shown in Table~\ref{tab:Clusterresults}, using the Feldman--Cousins approach~\cite{bib:FC}, incorporating detector related signal uncertainties in a semi-Bayesian approach~\cite{bib:Conrad}. 
%as well as the statistical uncertainty on the background evaluation. 
Astrophysical uncertainties are not included to simplify the inclusion of better estimates of the $J$ factors in the future. As an illustration of present uncertainties, in Fig.~\ref{fig:systematics} we show the impact of including the astrophysical uncertainty on the $J$ factor into the limit calculation for Segue~1.

%With the uncertainties on $J$ listed in table~\ref{tab:Dwarfsclusters}, the limits, obtained by by appropriately convoluting the lognormal uncertainty distribution for the $J$ factors, would soften by xx\%, yy\% and zz\% for the Segue 1 and Ursa Major II stacked results, M31, Virgo and Coma, respectively.

We present the upper limits for various objects and annihilation channels in the following plots. The sensitivity curves are the result of two competing
effects. One finds the effective area improves with increasing neutrino energies at higher WIMP masses,
%because the effective area increases with increasing neutrino energy,
while the background decreases. At the same high masses, the WIMP number density 
decreases, which ultimately reduces the WIMP annihilation rate.

\begin{figure}
  \includegraphics[width=1.05\columnwidth]{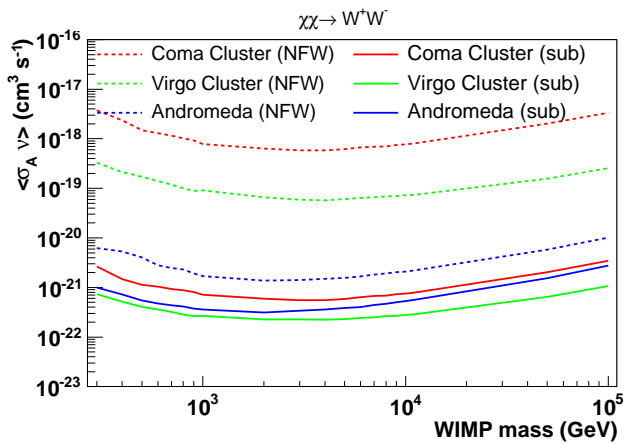}
  \caption{%Comparison of sensitivities (dashed lines) and upper limits (solid lines) 
    Upper limits for the Coma and Virgo clusters and the Andromeda galaxy for annihilation into $W^+W^-$. The dashed lines show the case for assumed pure NFW profiles, while the solid lines take into account substructures within the halos. %The corresponding sensitivities differ by constant factors of 
The sensitivity curves (not shown) differ from the limit curves by multiplicative factors of
    0.93 (Coma subhalos),
    0.99 (Coma NFW),
    1.1 (Virgo subhalos),
    1.2 (Virgo NFW),
    1.4 (Andromeda subhalos), and
    5.2 (Andromeda NFW).
  }
  \label{fig:subclusters}
\end{figure}

\begin{figure}[t]
  \includegraphics[width=1.05\columnwidth]{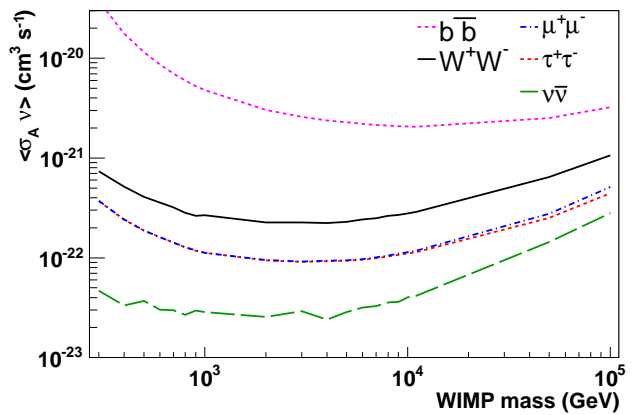}
  \caption{%Sensitivities (dashed lines) and upper limits (solid lines) 
    Upper limits for the Virgo cluster, correcting for subhalos, for annihilation into $b\bar{b}$, $W^+W^-$, $\tau^+\tau^-$, $\mu^+\mu^-$, and $\nu\bar{\nu}$. %The corresponding sensitivities differ by constant factors of 
The sensitivity curves (not shown) differ from the limit curves by multiplicative factors of
    1.3 ($\tau^{+}\tau^{-}$),
    0.98 ($b\bar{b}$),
    1.1 ($W^{+}W^{-}$),
    1.3 ($\mu^{+}\mu^{-}$), and
    1.3 ($\nu\bar{\nu}$).
  }
    \label{fig:virgodecay}
\end{figure}
\begin{figure}
\includegraphics[width=1.05\columnwidth]{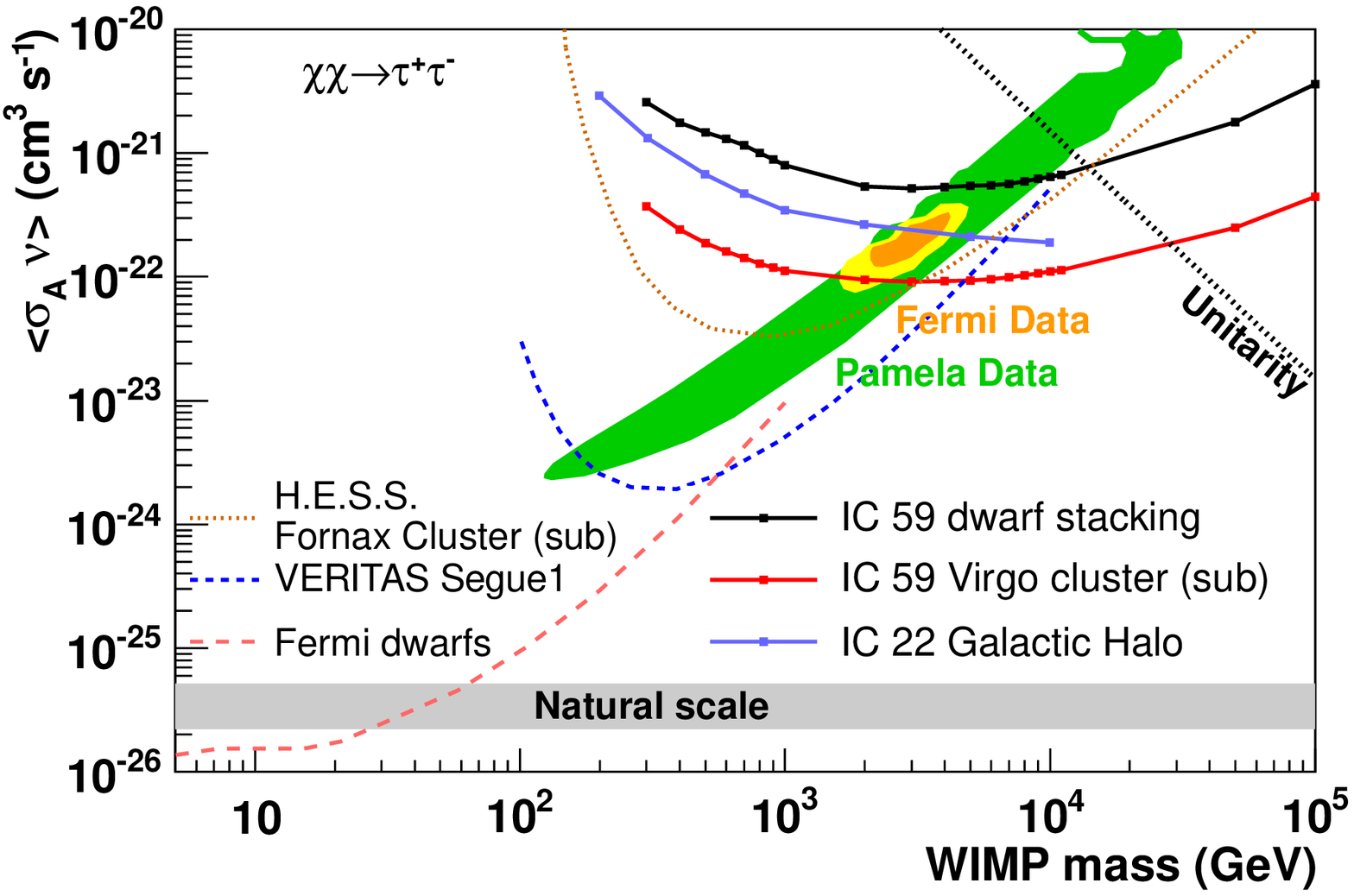}
\includegraphics[width=1.05\columnwidth]{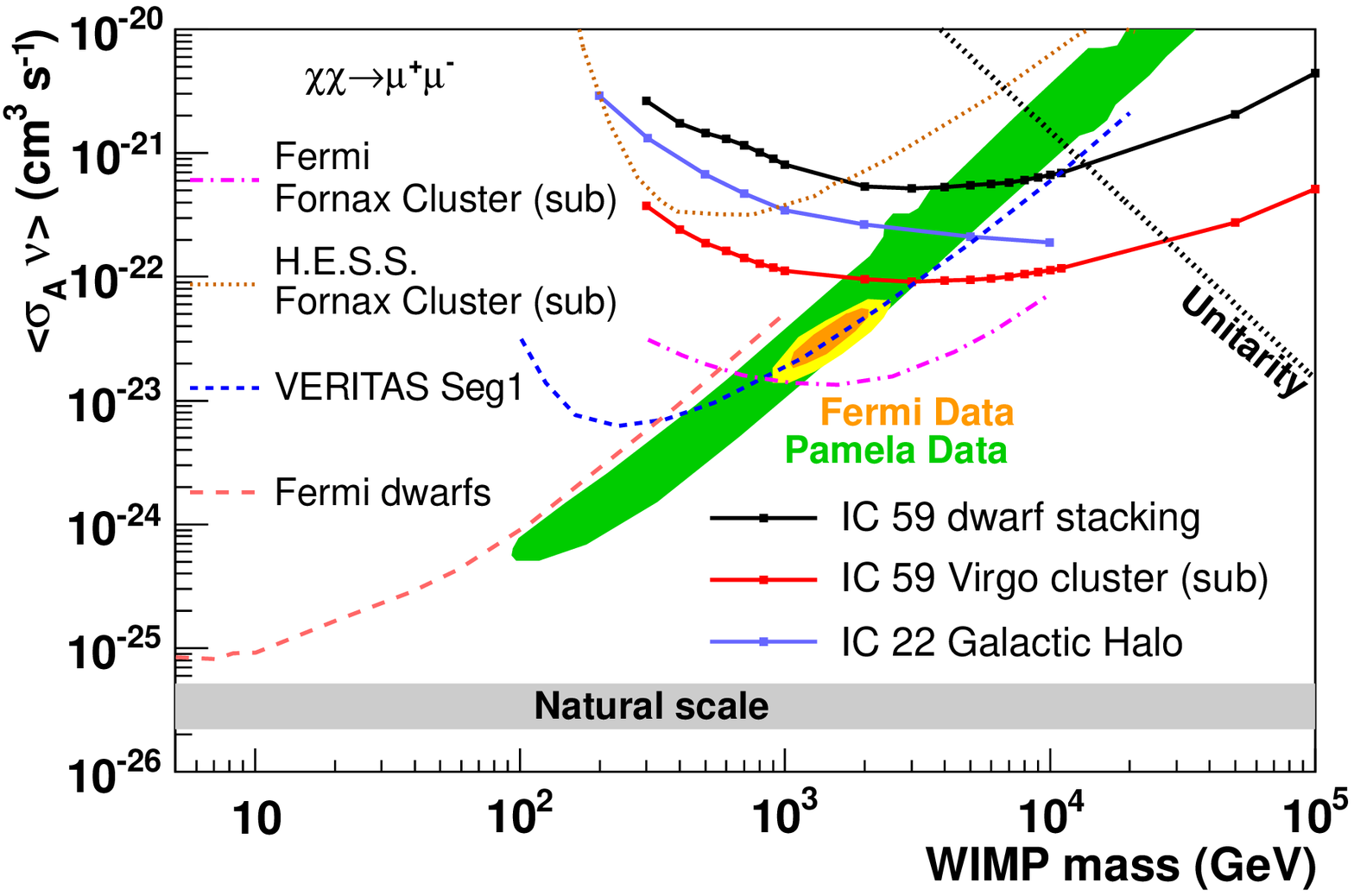}
	\caption{Summary of limits obtained in the $\tau^+\tau^-$ and $\mu^+\mu^-$ channels compared to preferred regions obtained by interpreting the PAMELA and Fermi excesses as due to dark matter annihilation~\cite{bib:Leptophilic2}. Also shown are limits from VERITAS for Segue~1~\cite{bib:veritas}, from H.E.S.S. for the Fornax galaxy cluster~\cite{bib:HESS}, from Fermi for stacked dwarf galaxies~\cite{bib:Jfactors} and the Fornax cluster~\cite{bib:FermiFornax}, as well as the IceCube result for the Galactic halo~\cite{bib:Halo}. Also shown is the ``natural scale'', for which the WIMP is a thermal relic~\cite{bib:Natural1,bib:Natural2} and the unitarity bound~\cite{bib:Unitarity1,bib:Unitarity2}, which limits the cross sections at high masses.}
	\label{fig:comparison_tau}
\end{figure}

Figure~\ref{fig:dwarf} compares the extracted upper limits for the dwarf galaxies assuming WIMP annihilation to the $W^+W^-$ channel. The best sensitivity is achieved for the stacked result of Segue~1 and  Ursa Major II. However, due to an underfluctuation of events, the most constraining limit for a single dwarf galaxy is obtained for Coma Berenices for WIMP masses above ~20 TeV. 
Figure~\ref{fig:subclusters} shows the effect of including boost factors due to subhalos. In this scenario the most stringent limit is achieved for the Virgo galaxy cluster, followed by Andromeda.  
Figur.~\ref{fig:virgodecay} compares the limits for the Virgo galaxy cluster (including subhalos) for each studied  annihilation channel. Because of the larger effective area of IceCube for higher energies, the most stringent limits are achieved for $\nu\bar{\nu}$ followed by the limits for $\tau^+\tau^-$, $\mu^+\mu^-$, and $W^+W^-$ channels.

Finally, in Fig.~\ref{fig:comparison_tau}, the limits for the $\tau^+\tau^-$ and $\mu^+\mu^-$ annihilation channels are compared to the preferred regions obtained by interpreting the PAMELA positron excess and electron data from Fermi and H.E.S.S. as being due to dark matter annihilation~\cite{bib:Leptophilic2}. The recent AMS-02 results will further tighten these regions. Included are results from $\gamma$-ray experiments and the ``natural cross section'' expected from the freeze-out of dark matter following production in the big bang~\cite{bib:Natural1,bib:Natural2}. The Fermi results strongly constrain the mass region below 1~TeV, while the results of IceCube provide valuable information for masses above. The limit from the Virgo galaxy cluster challenges the interpretation of the positron excess as being due to dark matter, %if WIMPs predominantly annihilate into channels providing hard neutrino spectra and if the boost factor is as large as predicted.  
if the boost factor is as large as predicted. The most stringent limits are achieved for annihilation channels providing hard neutrino spectra, which is complementary to searches by gamma telescopes.

\section{Summary}
\label{sect:Summary}
Using a sample of high-energy neutrinos collected during 2009--2010 with IceCube in its 59-string configuration, we have searched for a neutrino excess in the direction of the Virgo and Coma galaxy clusters, Andromeda (M31) as well as the Segue~1, Ursa Major II,  Coma Berenices, and Draco dwarf galaxies. Finding no significant excess, we placed constraints on the dark matter velocity averaged self-annihilation cross section, $\left< \sigma_A v \right> $,  at the 90\% C.L. for WIMP masses between 300~GeV and 100~TeV for a range of assumed WIMP annihilation channels. While $\gamma$-ray experimental observations provide significantly stronger limits below 1~TeV, our measurements competitively probe the cross section above 5 TeV in the $\chi  \chi \rightarrow \tau^+ \tau^-$ channel, particularly when incorporating the large effect of dark matter subhalos. Note that the tested cross sections are roughly a factor of 5000 above the natural scale, which can be accomplished by a substantial Sommerfeld enhancement~\cite{bib:sommerfeld1,bib:sommerfeld2}.
The results will improve in the future by incorporating more data from the fully instrumented IceCube detector and by employing a likelihood method for the stacking of potential sources. % and by analyzing the data for realistic mixtures of WIMP annihilation chan.\\

%{\em Outlook:}
%\begin{itemize}
%\item {\em improve stacking method} [Likelihood Methode]
%\item {\em include more data}
%item {\em use realistic mixture of decays} 
%\end{itemize}
%\appendix

\section{Acknowledgements}
\label{sect:Acknowledgements}
We acknowledge the support from the following agencies:
U.S. National Science Foundation-Office of Polar Programs,
U.S. National Science Foundation-Physics Division,
University of Wisconsin Alumni Research Foundation,
the Grid Laboratory Of Wisconsin (GLOW) grid infrastructure at the University of Wisconsin - Madison, the Open Science Grid (OSG) grid infrastructure;
U.S. Department of Energy, and National Energy Research Scientific Computing Center,
the Louisiana Optical Network Initiative (LONI) grid computing resources;
Natural Sciences and Engineering Research Council of Canada,
WestGrid and Compute/Calcul Canada;
Swedish Research Council,
Swedish Polar Research Secretariat,
Swedish National Infrastructure for Computing (SNIC),
and Knut and Alice Wallenberg Foundation, Sweden;
German Ministry for Education and Research (BMBF),
Deutsche Forschungsgemeinschaft (DFG),
Helmholtz Alliance for Astroparticle Physics (HAP),
Research Department of Plasmas with Complex Interactions (Bochum), Germany;
Fund for Scientific Research (FNRS-FWO),
FWO Odysseus programme,
Flanders Institute to encourage scientific and technological research in industry (IWT),
Belgian Federal Science Policy Office (Belspo);
University of Oxford, United Kingdom;
Marsden Fund, New Zealand;
Australian Research Council;
Japan Society for Promotion of Science (JSPS);
the Swiss National Science Foundation (SNSF), Switzerland;
National Research Foundation of Korea (NRF).
The research has made use of the SIMBAD database, operated at CDS, Strasbourg, France.

%% References


\bibliographystyle{model1a-num-names}
\bibliography{<your-bib-database>}



\end{document}